\documentclass[twocolumn]{aastex61}
\usepackage{amssymb}
\usepackage{graphicx}
\usepackage{natbib}

\begin{document}

\shorttitle{Accurate frequencies for S\lowercase{i}H$^+$}
\shortauthors{Dom\'enech, Schlemmer \& Asvany}
\submitjournal{The Astrophysical Journal}
\accepted{26/09/2017}

\title{Accurate frequency determination of vibration-rotation and rotational transitions of S\lowercase{i}H$^+$ }

\author[0000-0001-8629-2566]{Jos\'e L. Dom\'enech}
\affiliation{Instituto de Estructura de la Materia (IEM-CSIC), Serrano 123, E28006 Madrid, Spain}
\author[0000-0002-1421-7281]{Stephan Schlemmer}
\affiliation{I. Physikalisches Institut, Universit\"at zu K\"oln, Z\"ulpicher Str.~77,
50937 K\"oln, Germany}
\author[0000-0003-2995-0803]{Oskar Asvany}
\affiliation{I. Physikalisches Institut, Universit\"at zu K\"oln, Z\"ulpicher Str.~77,
50937 K\"oln, Germany}
\correspondingauthor{Jos\'{e} L. Dom\'{e}nech}
\email{jl.domenech@csic.es}

\begin{abstract}
The fundamental $^{28}$SiH$^+$ ion has been characterized in a collaborative work, utilizing a hollow-cathode-discharge laser-spectrometer
and a cryogenic ion trap spectrometer. Twenty-three vibration-rotation transitions around 4.75 $\mu$m 
have been detected with high accuracy. This has facilitated the first direct measurement of the pure rotational 
transition $J=1\leftarrow0$ at 453056.3632(4)~MHz in the trap spectrometer. The measured and accurately predicted 
transitions enable the search for this ion in space with IR and sub-mm telescopes.
\end{abstract}


\newcommand{\wn}{cm$^{-1}$}
\newcommand{\sihp}{SiH$^+$}

\section{Introduction}

About twelve of the molecules identified in the interstellar medium contain silicon, mostly bound to C, N, O, or S atoms. On the other hand, silane (SiH$_4$)  is the
only one containing hydrogen that has been securely identified (by \citet{Goldhaber1984} in the outer envelope of IRC+10216).
Merely tentative detections have been reported for SiH$_2$ \citep{Turner1995}, 
SiH \citep{Schilke2001} and SiH$_3$CN \citep{Agundez2014}.  
The simplest silicon hydrides, the radical SiH (sylilydyne) and the ion SiH$^+$ (sylylidynium)  
have only been identified in the sun's photosphere \citep{Sauval1969,Grevesse1970}, 
and SiH also in the atmosphere of some cool stars (see e.g. \citet{Merrill1953,Merrill1959}).

In diffuse and translucent clouds, hydrides can usually be observed by their electronic absorptions at visible and ultraviolet wavelengths superposed on the spectra of bright
background stars.  Unlike the chemically similar CH$^+$, this is not the case for \sihp.
At the time of the discovery of SiH$^+$ in the sun, it was expected to be found in the ISM on the grounds of the relatively high abundance of Si, the similarity to CH$^+$, and the 
ionization potential of Si being lower than that of C, that would allow Si to remain ionized to larger optical depths \citep{Douglas1970,DeAlmeida1978}. 
The gas phase chemistry of Si in diffuse clouds  was discussed by \citet{Turner}, who pointed out radiative association as the main formation process, 
and dissociative recombination with electrons, photodissociation 
and reactions with H and O as destruction mechanisms.  Unlike CH$^+$, H$_2$ does not destroy SiH$^+$, and that could favor a higher \sihp/CH$^+$ abundance ratio than that of Si/C.
Since radiative association of Si$^+$ with H and H$_2$ is very slow \citep{Stancil2000}, in cold dense clouds, the ion-molecule silicon chemistry is initiated by the exothermic 
reaction H$^+_3$ + Si $\rightarrow$ SiH$^+$ + H$_2$.  The reaction of \sihp\ with H$_2$ is endothermic, so the main destruction pathway for sylylidynium is mainly collisions 
with C, N, O \citep{Herbst1989}.  In dark clouds most gas-phase Si is supposed to be in the form of SiO, for which extremely low upper limits have been derived and for which 
models predict a much higher abundance, leading to the conclusion that Si has to be very heavily depleted into grains. \citet{McKay1996} has pointed out that depletion would 
not need to be so high if SiH$_4$ were the major Si-bearing species. The formation reaction ${\rm Si^+ + H_2 \rightarrow SiH^+ + H}$ is highly endothermic 
(by $\sim$14200 K, from the thermochemical data of \citet{Curtiss1988}), therefore,   and similarly to the formation mechanisms suggested for the ubiquitous 
CH$^+$, or SH$^+$, a warm chemistry is necessary, and this reaction would only proceed in shocked or turbulent regions, or if hydrogen molecules were vibrationally 
excited ($v\!\!\geqslant\!\!3$). It is worth noting that sulfanylium (the SH$^+$ radical) was searched for, and not found, in absorption in the UV \citep{Millar1988,Magnani1991}, 
until it was recently identified in the sub-mm range by \citet{Menten2011}.  The endothermicity of the reaction ${\rm S^+ + H_2 \rightarrow SH^+ + H}$ 
is 9860 K, more than twice of that involving CH$^+$ (4280 K), and vibrationally excited H$_2$ (in $v>1$ and $v\geqslant2$, respectively) has been invoked 
in both cases as a way to overcome the energy barriers \citep{Gerin2016, Zanchet2013a}.

Laboratory spectroscopic information on SiH$^+$ is surprisingly scarce.  The electronic spectrum in the vis-UV region ($A^1\Pi-X^1\Sigma^+$ transition) 
was recorded by \citet{Douglas1970}, leading to its identification in the sun's photosphere \citep{Grevesse1970}.  Based on those data, \citet{Singh1978} 
estimated line positions for rotation and vibration-rotation lines of SiH$^+$.  In particular, the predicted frequency for the $J=1-0$ transition 
in $v=0$ was 453037~MHz, with no estimation of the uncertainty.  \citet{Carlson1980} identified two new bands of the electronic transition, and improved the molecular constants.
\citet{Davies1988} detected the fundamental vibration-rotation band of SiH$^+$ in a silane plasma with diode laser spectroscopy.  
Most likely due to the patchy coverage of lead-salt diode lasers, they could measure only seven lines in the $R$-branch and just one 
in the $P$-branch, with a quoted accuracy of $\sim 0.003$~\wn.  From a fit to their data we obtain 453121(40)~MHz  for the $J=1-0$ rotational line.  
Until now there was no direct measurement of the rotational transitions.  Regarding transition dipole moments, \citet{Singh1978} 
estimated the transition dipole moment for the $v=1-0$ band to be $\mu_{1-0}=0.1081$~D, and the permanent electric dipole moment 
has been calculated by \citet{Park1992} to be $\mu_0=0.5314$~D , while \citet{Muller2013} give $\mu_0=0.388-0.454$~D with higher level calculations.

In this work we have measured 23 vibration-rotation lines of the $v=1-0$ band of \sihp\ with high accuracy ($\sim$ 10 MHz, $3\sigma$), 
using a difference frequency laser spectrometer coupled to a hollow cathode discharge. Besides providing a more accurate and 
complete set of infrared frequencies than previously available, it has allowed for an accurate prediction of the $J=0-1$ rotational transition.  
The frequency of the latter has been  measured with sub-ppb accuracy in a cold ion trap using the method of state-dependent attachment of helium atoms.
\clearpage

\section{Experimental setup and results}
\begin{figure}
\caption{\label{figure:IRlines} Two examples of detected lines of \sihp\ and their Gaussian fits.  
Both lines are the result of averaging 100 scans (1000 s integration time.)}
\vspace{10pt}
\includegraphics[width=0.95\columnwidth]{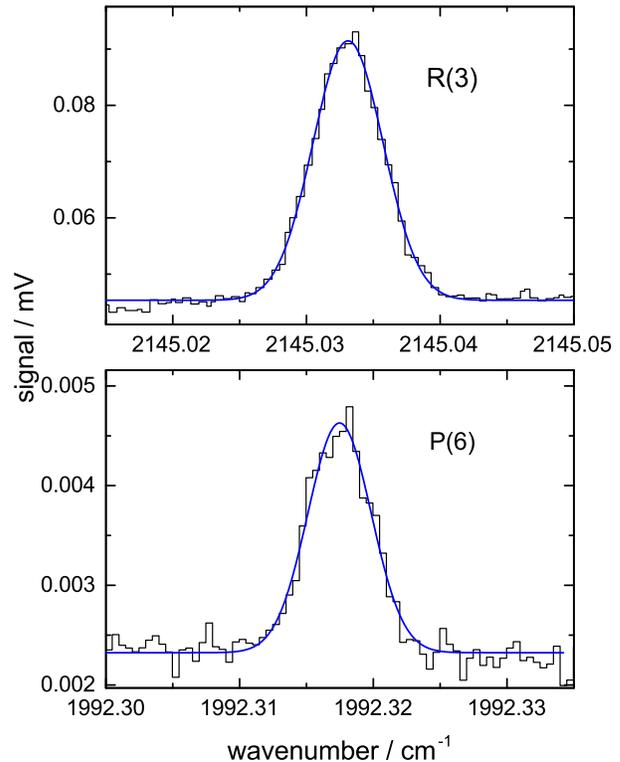}
\end{figure}

The vibration-rotation spectrum was measured in Madrid with a difference frequency laser spectrometer coupled to a hollow cathode discharge,
using a double modulation of the discharge current (i.e.\ concentration modulation) and the infrared beam power.  This setup has been used recently
for the measurement of the $\nu_4$ band of the NH$_3$D$^+$ ion \citep{Cernicharo2013,Domenech2013}, 
the measurement of the fundamental vibration-rotation bands of  $^{36}$ArH$^+$, $^{38}$ArH$^+$ \citep{Cueto2014}, and
H$^{35}$Cl$^+$ and H$^{37}$Cl$^+$ \citep{Domenech2016a}.  Since it has already been described in those works, only the details relevant to the
present experiment will be given here. The accuracy of the frequency scale relies on an Ar$^+$ laser which is frequency-locked to a
hyperfine transition of $^{127}$I$_2$.  The same laser is also used to calibrate a high accuracy wavemeter (10~MHz, $3\sigma$) that
measures the frequency of the dye laser at each data point, leading to an absolute accuracy of the IR frequencies limited by the
wavemeter (10 MHz), and an internal precision on the order of 1~MHz. The precursor gases were silane and hydrogen.  Since the pyrophoric
nature of silane poses major difficulties for its handling, we decided to use a premixed sample of 0.5\% SiH$_4$ in He
(Alphagaz $(0.50\pm0.01)\%$ SiH$_4$ UHP in He N46).  Adding H$_2$ to the flowing mixture was found to increase SiH$^+$ signals. To our surprise,
we found that, after some hours of operation, the SiH$_4$/He was not necessary at all, and that the signal level was maintained for rather long
periods with only H$_2$ in the discharge.  
Since a deposit of amorphous Si was forming on the cathode walls, this suggested that electron impact was removing Si atoms from the cathode and
that the reaction Si+H$_3^+\rightarrow$ SiH$^+$+ H$_2$ was being very efficient in the production of SiH$^+$.  This was in agreement with the
observation of \citet{Davies1988} although they did not observe such a marked persistence of the signal.  Finally, the experiments were
run at 1~mbar total pressure with 0.5 mbar each of H$_2$ and SiH$_4$/He mixture, and at room temperature.  The discharge was maintained at
375~mA and 800~V.   In the present experiment the discharge  modulation frequency could be increased up to 15~kHz, and it was found
that $\sim$10~kHz provided a good compromise beween signal intensity and discrimination of ion signals against interferences of other
species (neutral SiH$_4$ and likely other polysilanes).
A total of 23 lines was measured, from $P(11)$ to $R(11)$, spanning the range 1902-2236~\wn.  Between 100 and 500 scans were averaged for each
line, reaching signal to noise ratios between 100 and 10, depending on the signal intensity.  Two examples of the detected lines are shown in
Figure~\ref{figure:IRlines}.  The initial search was facilitated by the wavenumbers listed in \citet{Davies1988}, although, as the number of detected
lines increased, our own data were used to predict the next lines.  Gaussian fits to the line profiles were used to extract the line centers.
Table~\ref{tab1} lists the observed wavenumbers, their 1$\sigma$ uncertainty (calculated as the quadratic sum of the center frequency uncertainty
derived from the fit and the 1$\sigma$ accuracy of the wavemeter) and the comparison with the measured frequencies of \citet{Davies1988}.
Our frequencies were fit using the program PGOPHER \citep{Wes17} for the case of a linear molecule in a $^1\Sigma^+$ state.  Up to the sextic
centrifugal distortion constants could be derived with statistical significance, both for $v$=0 and $v$=1.
For the $J$=1--0 fundamental rotational transition, a frequency prediction of 453057.7 MHz, with $3\sigma$=1.5~MHz, was obtained.

\begin{table}[h]  
\caption{\label{tab1} Wavenumbers of vibration-rotation (in \wn) and frequencies of pure rotational transitions  (in MHz) of  $^{28}$SiH$^+$.  
The final error is given in parentheses in units of the last digit. The state-dependent attachment method was not able to 
detect the  $2  \leftarrow  1$ rotational transition, therefore only a prediction is given. For further laboratory and astronomical searches,
also the predictions for the next two rotational transitions are listed.}
 \begin{center}
 \scriptsize
\begin{tabular}{rclr@{}ll}
\hline
$(v',J')$&$\leftarrow$&$(v'',J'')$ &  \multicolumn{2}{c}{This work} & Former work$^a$       \\
\hline
    (1,10)&$\leftarrow$&(0,11) & \multicolumn{2}{c}{1902.42994(25)} &  \\
    (1,9) &$\leftarrow$&(0,10) & \multicolumn{2}{c}{1921.04096(16)} &  \\
    (1,8) &$\leftarrow$& (0,9) & \multicolumn{2}{c}{1939.34400(11)} &  \\
    (1,7) &$\leftarrow$& (0,8) & \multicolumn{2}{c}{1957.33110(14)} &  \\
    (1,6) &$\leftarrow$& (0,7) & \multicolumn{2}{c}{1974.99216(12)} &  \\
    (1,5) &$\leftarrow$& (0,6) & \multicolumn{2}{c}{1992.31748(11)} &  \\
    (1,4) &$\leftarrow$& (0,5) & \multicolumn{2}{c}{2009.29756(10)} &  \\
    (1,3) &$\leftarrow$& (0,4) & \multicolumn{2}{c}{2025.92322(10)} &  2025.923\\
    (1,2) &$\leftarrow$& (0,3) & \multicolumn{2}{c}{2042.18438(10)} &  \\
    (1,1) &$\leftarrow$& (0,2) & \multicolumn{2}{c}{2058.07203(11)} &  \\
    (1,0) &$\leftarrow$& (0,1) & \multicolumn{2}{c}{2073.57690(10)} &  \\
    (1,1) &$\leftarrow$& (0,0) & \multicolumn{2}{c}{2103.40022(11)} &  \\
    (1,2) &$\leftarrow$& (0,1) & \multicolumn{2}{c}{2117.70038(10)} &  \\
    (1,3) &$\leftarrow$& (0,2) & \multicolumn{2}{c}{2131.58102(11)} &  \\
    (1,4) &$\leftarrow$& (0,3) & \multicolumn{2}{c}{2145.03311(11)} &  \\
    (1,5) &$\leftarrow$& (0,4) & \multicolumn{2}{c}{2158.04819(10)} &  2158.055\\
    (1,6) &$\leftarrow$& (0,5) & \multicolumn{2}{c}{2170.61766(11)} &  2170.61 \\
    (1,7) &$\leftarrow$& (0,6) & \multicolumn{2}{c}{2182.73306(12)} &  2182.737\\
    (1,8) &$\leftarrow$& (0,7) & \multicolumn{2}{c}{2194.38565(11)} &  2194.387\\
    (1,9) &$\leftarrow$& (0,8) & \multicolumn{2}{c}{2205.56783(11)} &  2205.573\\
    (1,10)&$\leftarrow$& (0,9) & \multicolumn{2}{c}{2216.27186(11)} &  2216.273\\
    (1,11)&$\leftarrow$&(0,10) & \multicolumn{2}{c}{2226.48921(25)} &  2226.492\\
    (1,12)&$\leftarrow$&(0,11) & \multicolumn{2}{c}{2236.21237(61)} &  \\
\hline
    (0,1)&$\leftarrow$&  (0,0) &  453056&.3632(4)     &   \\
    (0,2)&$\leftarrow$&  (0,1) &  905837&.055(55)$^b$ &   \\
    (0,3)&$\leftarrow$&  (0,2) & 1358066&.73(21)$^b$ &   \\
    (0,4)&$\leftarrow$&  (0,3) & 1809470&.70(47)$^b$ &  \\
\hline
\end{tabular}
\end{center}
$^a$\cite{Davies1988}. The uncertainty is 0.003~\wn.\\
$^b$ Predicted\\
\end{table}

\begin{figure}
\includegraphics[width=\columnwidth]{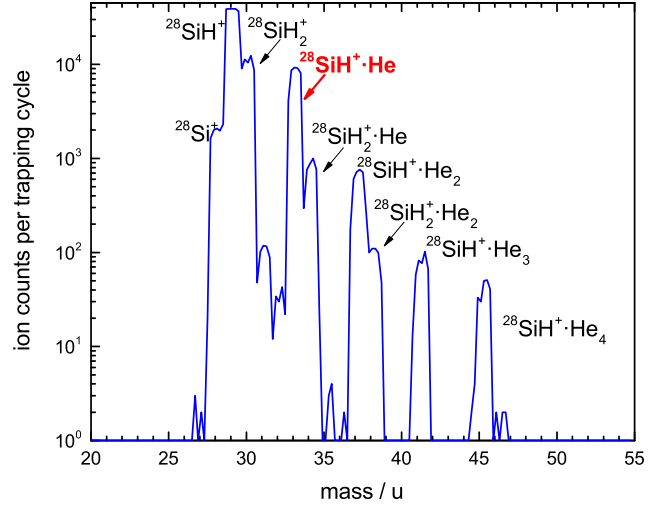}
\caption{\label{fig2} Mass spectrum recorded after the SiH$^+$ ions have been trapped for 700~ms in the 4~K ion
  trap filled with Helium  ($n \approx 10^{14}$~cm$^{-3}$).
Attachment of up to four He atoms is observed. The ion signal  SiH$^+\cdotp$He ($m=33$~u) has been used for rotational spectroscopy, see Fig.~\ref{fig3}.
Due to imperfect mass selection before the trapping cycle, also some traces of  Si$^+$ and  SiH$_2^+$ are admitted to the trap.}
\end{figure}

\begin{figure}
\includegraphics[width=0.95\columnwidth]{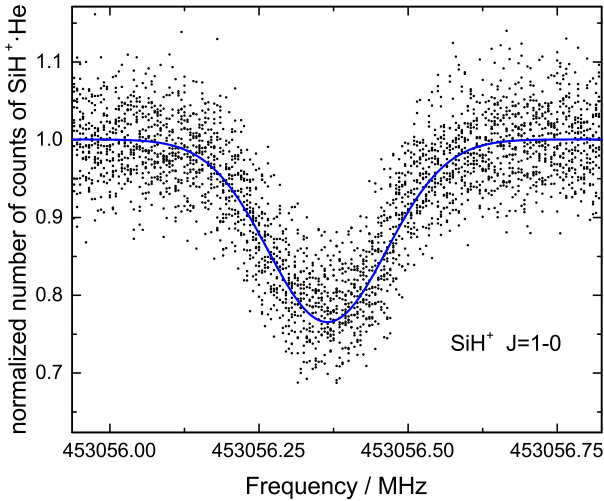}
\caption{\label{fig3} Normalized signal of the $J=1 \leftarrow 0$ transition of \sihp, measured as depletion
of the SiH$^+\cdotp$He counts. The fit of this measurement with the shown Gaussian 
yields the center frequency 453056.3664(15)~MHz. With eleven of such measurements 
the final value given in Table~\ref{tab1} is obtained}.
\end{figure}

The fundamental rotational transition of SiH$^+$ was then measured in the K\"oln 
laboratories  exploiting the rotational state dependence of 
the attachment of He atoms to cations (\cite{bru14,bru17,jus17}). 
The experiment was performed in the 4~K trapping machine COLTRAP described by \cite{asv10,asv14}.
The $^{28}$SiH$^+$ ions were generated in a storage ion source by 
bombarding the SiH$_4$ precursor gas (Linde AG, 2\% silane in He) with 30~eV electrons.
A pulse of several ten thousand mass-selected ions (m=29~u) was injected into 
the 22-pole ion trap  filled with about 10$^{14}$ cm$^{-3}$ He.
During the trapping time of  700~ms, the complexes SiH$^+\cdotp$(He)$_n$ ($n=1-4$) formed 
by three-body collisions, as shown in Fig.~\ref{fig2}. The detection of the resonant absorption of the sub-mm wave by the naked ion ($n=0$) 
was achieved by observing the decrease of the counts of the complex with $n=1$ (mass 33~u).   
The sub-mm wave radiation was supplied by an atomic clock referenced synthesizer (Agilent E8257D) driving a  multiplier chain source
(Virginia Diodes, Inc.) covering the range 80-1100~GHz.
The nominal sub-mm power of the source at the fundamental rotational transition of SiH$^+$ was about 160~$\mu$W. 
This power was attenuated by a factor of about three  (with further losses by the transport to the trap) 
in order to avoid power broadening. In this case, the spectral profile is a pure Gaussian with a corresponding 
Doppler temperature $T=13$~K, as seen in Fig.~\ref{fig3}. The final value for the fundamental  $J= 1  \leftarrow  0$ after 
eleven measurements is 453056.3632(4) MHz (1$\sigma$) and is also listed in Table~\ref{tab1}.
The resulting uncertainty of 400~Hz correponds to a relative precision of better than 1~ppb.
Also, the very good agreement with the prediction from the IR--only data is notheworthy. 
Interestingly, we were not able to detect the  $J= 2  \leftarrow  1$ transition, even though the corresponding
line for e.g.\ CD$^+$ has been observed with the same method. 
The joint effect of the small dipole moment for SiH$^+$, lower available power, small population in $J=1$,
and, most probably, a small difference in He attachment rate coefficients for $J=1$ and $J=2$ can explain this non-observation.
Therefore, only a prediction based on all our data is given in Table~\ref{tab1}. 


We also searched for the fundamental transitions of isotopologues of \sihp. Although $^{30}$Si is less abundant 
than $^{29}$Si (3.1~\% and 4.7~\%, respectively), we choose it for the search scan because it has nuclear 
spin $I=0$ and therefore a potential signal dilution due to hyperfine effects can be excluded.
Our predicted frequency value (based on the mass scaling of the spectroscopic parameters) was about 452014~MHz.
We integrated in the range 451970 - 452030~MHz, and did not find any rotational signal of $^{30}$SiH$^+$.
This non-detection is caused by the probed ion mass ($m=35$~u) 
being  heavily dominated by $^{28}$SiH$_3^+\cdotp$He, with the desired $^{30}$SiH$^+\cdotp$He
ions making up much less than 3.1~\% on that mass.

%


\section{Spectroscopic parameters}
The measured frequency of the pure rotational line $J=1\leftarrow0$ was added to the fit, rendering the final set of
constants given in Table \ref{tab2}. The frequency predictions for the 
higher rotational transitions in Table \ref{tab1} have been calculated with these constants.
\begin{table}[h]
\caption{\label{tab2} Derived spectroscopic parameters (in \wn) from a fit to all measured transitions.
    Numbers is parentheses are one standard deviation in units of the last digit.}
\begin{center}
\begin{tabular}{lr@{}lr@{}l}
\hline
Parameter            & \multicolumn{2}{c}{$v=0$}       & \multicolumn{2}{c}{$v=1$}  \\
\hline
  $\nu$              &      &                     &  2088&.689199(67) \\    
  $B_v$              &     7&.556933267(160)      &     7&.35616891(602) \\ 
  $D_v\times10^4$    &     3&.832640(806)         &     3&.78803(123)    \\ 
  $H_v\times10^8$    &     1&.5188(595)           &     1&.3620(685)     \\ 
\hline
\end{tabular}
\end{center}
\end{table}

\section{Conclusion and Outlook}

We have measured  23 lines of the fundamental vibration-rotation band of $^{28}$SiH$^+$, with an absolute 
accuracy better than 10 MHz (3.3$\times10^{-4}$ \wn), fifteen of these 
for the first time.  This has facilitated the first direct and very accurate measurement of the pure rotational 
transition $J=1\leftarrow0$ at 453056.3632(4)~MHz, and a prediction for the $J=2\leftarrow1$ transition at 905837.055(60)~MHz.
It is gratifying to see how well high resolution IR spectroscopy can guide laboratory (and astronomical) searches for rotational transitions.

Although current astrochemical models do not predict a high abundance of SiH$^+$ (see e.g.\ \cite{Herbst1989}), 
the importance of the molecule together with the ever increasing sensitivity and resolution of radio and IR telescopes, 
warrant the interest of providing precise values for the transition frequencies.  

The $J=0-1$ transition was outside the bands covered by {\it Herschel}.  An examination of {\it Herschel's} data towards 
SgrB2 and Orion has not revealed any signature that could be attributed to the $J=2-1$ line.  Now that the {\it Herschel} mission has ended, 
other alternatives may be ALMA, APEX or SOFIA.  The $J=1-0$ line lies within ALMA band 8, and within the range covered by the instrument APEX-3, but it is close 
to a telluric water line.  With a total precipitable water vapor (pwv) of 0.5 mm, the calculated transmission of the atmosphere at ALMA site is 0.38, while for the average 1 mm pwv it drops to 0.15.  The $J=2-1$ line lies within ALMA band 10, 
and the instrument CHAMP+ at APEX, although also with limited atmospheric transparency (calculated transmission is 0.28 and 0.07 for 0.5 mm and 1 mm pwv, respectively).  Observations of the $J=0-1$ transition  could be attempted in dry weather with both ALMA or APEX if enough integration time was available.  However, given the not so favorable atmospheric conditions, and the 
larger Einstein coefficients of the vibration-rotation transitions, observations in the mid-infrared (around 4.75~$\mu$m) could also be a good alternative to detect this molecule. 
The instruments EXES onboard SOFIA, iSHELL at IRTF, or the future CRIRES+ at VLT could offer good opportunities to detect absorptions by this molecule in the ISM.
The lines of sight would be those of the diffuse medium, shock regions or supernova ejecta against bright IR sources. 

\acknowledgments
This work (including the research stay of J.L.D in K\"oln) has been  supported by the Deutsche
Forschungsgemeinschaft (DFG) via SFB 956 project B2
and the Ger\"atezentrum "Cologne Center for Terahertz Spectroscopy".
J.L.D acknowledges partial financial support from the Spanish MINECO through grant
FIS2016-77726-C3-1-P and from the European Research Council through grant
agreement ERC-2013-SyG-610256-NANOCOSMOS.
We thank Holger M\"uller for stimulating discussions. 
We also thank P. Schilke, J. Cernicharo and V. Ossenkopf-Okada 
for checking their observations made with {\it Herschel} for the \sihp\ $2-1$ line.
The authors gratefully acknowledge the work done over the last years 
by the electrical and mechanical workshops of the I. Physi\-kali\-sches Institut, 
as well as by the technicians at the Molecular Physics department of IEM-CSIC and CFMAC-CSIC workshops.

\software{PGOPHER \citep{Wes17}}

%

\end{document}